\documentstyle[seceq,preprint,psfig]{ptptex}

\preprintnumber[3cm]{
SMUHEP/99-03}

\inst{Division of Phisics and Mathematics, Setsunan University, Osaka 572-8508
\\
$^*$Department of Physics, Southern Methodist University, Dallas TX 75275 }

\abst{We consider constructing a canonical quantum theory of the light-cone 
gauge ($A_-$=0) Schwinger model in the light-cone representation. 
Quantization conditions are obtained by requiring that translational generators
$P_+$ and $P_-$ give rise to Heisenberg equations which, in a physical 
subspace, are consistant with the field equations. A consistent operator 
solution with residual gauge degrees of freedom 
is obtained by solving initial value problems on the light-cones.  The 
construction allows a parton picture although we have a physical vacuum with 
nontrivial degeneracies in the theory. }

\begin{document}

\title{Light-Cone Quantization of the Schwinger Model }
\author{Yuji {\sc Nakawaki} and Gary {\sc McCartor}$^{*}$}
\maketitle

\section{Introduction}

We will use the following notation:

\begin{eqnarray*}
&g^{++}=g^{--}=0,\quad g^{+-}=g^{-+}=2,\quad g_{++}=g_{--}=0,\quad
g_{+-}=g_{-+}={\frac{1}{2}},& \\
&x^{+}=x^{0}+x^{1},\quad x^{-}=x^{0}-x^{1},\quad {\partial }_{+}={\frac{1}{2}%
}({\partial }_{0}+{\partial }_{1}),\quad {\partial }_{-}={\frac{1}{2}}({%
\partial }_{0}-{\partial }_{1}),& \\
&{\gamma }^{0}={\sigma }_{1},\quad {\gamma }^{1}=i{\sigma }_{2},\quad {%
\gamma }^{5}=-{\sigma }_{3}& \\
&\Psi =\left( 
\begin{array}{c}
\Psi _{-} \\ 
\Psi _{+}
\end{array}
\right) \quad m={\frac{e}{\sqrt{\pi }}}.&
\end{eqnarray*}

Previously, one of us (Nakawaki)$^{1)}$ gave the following operator solution
to the Schwinger model in Landau gauge: 
\begin{equation}
\Psi _{+}=Z{\rm e}^{\Lambda _{+}^{(-)}}\sigma _{+}^{s}{\rm e}^{\Lambda _{+}^
{(+)}}
\end{equation}
\begin{equation}
\Lambda _{+}=-i\sqrt{\pi }(2\phi _{+}(x^{-})+\tilde{\eta}+\tilde{\Sigma}%
(x^{+},x^{-}))
\end{equation}
\begin{equation}
Z^{2}={\rm e}^{\gamma }\sqrt{{\frac{m\kappa }{8\pi ^{2}}}}
\end{equation}
\begin{equation}
\Psi _{-}=Z{\rm e}^{\Lambda _{-}^{(-)}}\sigma _{-}^{s}{\rm e}^{\Lambda _{-}^
{(+)}} \label{pmcl}
\end{equation}
\begin{equation}
\Lambda _{-}=-i\sqrt{\pi }(2\phi _{-}(x^{+})-\tilde{\eta}-\tilde{\Sigma}%
(x^{+},x^{-}))
\end{equation}
\begin{equation}
A_{\mu }=-m^{-1}\varepsilon _{\mu \nu }\partial ^{\nu }(\tilde{\eta}+\tilde{%
\Sigma}).
\end{equation}
Here, $\phi _{+}(x^{-})$ is the right moving component of a
Klaiber$^{2)}$-regulated (with parameter $\kappa$), free, massless scalar field
composed of the fusion operators: 
\begin{equation}
\phi _{+}^{(+)}(x^{-})=i(4\pi )^{-{\frac{1}{2}}}\int_{-\infty
}^{0}dk_{1}k_{0}^{-1}c(k_{1})\left({\rm e}^{-ik\cdot x}-\theta (\kappa
-k_{0})\right)
\end{equation}
\begin{equation}
\phi_+^{(-)} = (\phi_+^{(+)})^*.
\end{equation}

Where $c(k_{1})$ are the fusion operators associated with Bosonizing$^{1)}$ the
Fermi field$^{1)}$ (they satisfy the usual Boson commutation relations). $%
\phi _{-}(x^{+})$ is the left moving component of that field: 
\begin{equation}
\phi _{-}^{(+)}(x^{+})=i(4\pi )^{-{\frac{1}{2}}}\int_{0}^{\infty
}dk_{1}k_{0}^{-1}c(k_{1})\left({\rm e}^{-ik\cdot x}-\theta 
(\kappa -k_{0})\right)
\end{equation}
\begin{equation}
\phi_-^{(-)} = (\phi_-^{(+)})^*
\end{equation}
$\tilde{\eta}$ is a psuedoscalar ghost field given by: 
\begin{equation}
\tilde{\eta}^{(+)}=i(4\pi )^{-{\frac{1}{2}}}\int_{-{\infty}}^{\infty}dk_{1}k_{1}^{-1}\eta
(k_{1})\left( {\rm e}^{-ik\cdot x}-\theta (\kappa -k_{0})\right),  \label{epm}
\end{equation}
where: 
\begin{equation}
\lbrack \eta (k_{1}),\eta ^{\ast }(q_{1})]=-k_{0}\delta (k_{1}-q_{1})
\end{equation}
The field, $\tilde{\Sigma}$ --- the field associated with the physical
Schwinger particles --- is a massive psuedoscalar field of mass $m$
 ($={\frac{e}{\sqrt{\pi }}}$). The spurions are given in terms of these modes 
as: 
\begin{eqnarray}
&\sigma _{+}^{s}={\rm exp}\Big[i\sqrt{\pi }\{Q_{5}+Q\}(4m)^{-1}& \nonumber \\
&+2^{-1}\int_{-\kappa}^{\kappa }dk_{1}k_{1}^{-1}\{\eta (k_{1})-\eta ^{\ast
}(k_{1})\}+\int_{-\kappa }^{0}dk_{1}k_{0}^{-1}\{c(k_{1})-c^{\ast }(k_{1})\}
\Big]&
\end{eqnarray}
\begin{eqnarray}
&\sigma _{-}^{s}={\rm exp}\Big[i\sqrt{\pi }\{Q_{5}-Q\}(4m)^{-1}& \nonumber \\
&-2^{-1}\int_{-\kappa}^{\kappa }dk_{1}k_{1}^{-1}\{\eta (k_{1})-\eta ^{\ast
}(k_{1})\}+\int_{0}^{\kappa }dk_{1}k_{0}^{-1}\{c(k_{1})-c^{\ast }(k_{1})\}%
\Big].&
\end{eqnarray}
To be invariant under the large gauge transformations the vacuum must be
chosen to be a theta-state, formed as: 
\begin{equation}
|\Omega (\theta )\rangle \equiv \sum_{n=-\infty }^{\infty }{\rm e}^{iM\theta
}|\Omega (M)\rangle \quad ;\quad |\Omega (M)\rangle =({\sigma }_{+}^{\ast }{%
\sigma }_{-})^{M}|0\rangle.
\end{equation}
The physical subspace is formed by applying all polynomials in $\tilde{\Sigma%
}$ to $|\Omega (\theta )\rangle $.

Further details, and proof that the construction is an operator solution can
be found in the reference. Here we wish to point out that while the above
solution is representation independent, it is straightforward to formulate
the problem as an initial value problem at $t=0$, and thus find a solution in
the equal-time representation. Gauge invariant point splitting (in a
space-like direction) provides the necessary regulation for the operator
products. If we contemplate the initial value problem on the characteristic
surfaces, $x^{+}=0$ and $x^{-}=0$, things are not so straightforward. The
problem is that the Fermi products cannot be regulated by splitting in a
light-like direction. The $\tilde{\eta}$ field is the sum of a function of $%
x^{+}$ and a function of $x^{-}$, so the operator products are not regulated
by splitting in either light-like direction. Also, the $\tilde{\Sigma}$
field suffers an apparent, but spurious, infrared singularity due to the
fact that its massive character is not manifest at light-like separations.
Note that these problems only have to do with the formulation of the theory
as an initial value problem on the characteristics; the light-cone
representation perfectly well exists --- that is, modes of the fields along
the characteristics provide all the operators necessary to generate the
entire representation space. We do need modes along both characteristics but
all the physical operators except those necessary to define the vacuum can
be generated using modes along $x^{+}=0$. To generate the vacuum we need
both spurions and thus modes from both characteristics. We shall return to
the problem of formulating an initial value problem on the characteristic
surfaces below, but first we wish to consider the question of light-cone
gauge.

\section{Light-cone gauge}

We may attempt to reach the light-cone gauge by performing a nonlocal gauge
transformation on the Landau gauge solution. If we use the gauge function: 
\begin{equation}
\Theta = m^{-1}(\tilde{\eta} + \tilde{\Sigma})
\end{equation}
we find that $A^+ = 0$. The resulting construction almost works but it is
not quite right. One problem is that the $x^-$-dependent parts of the $%
\tilde{\eta}$ and ${\phi}$ fields are not natural degrees of freedom in
light-cone gauge. That is, standard quantization methods, whether in the
equal-time representation or in the light-cone representation do not include
those degrees of freedom in the representation space. Those degrees of
freedom decouple, at least formally, and we shall simply remove them from
the solution. The other problem is that we have performed the gauge
transformation with the Klaiber-regulated $\tilde{\eta}$ field but have left
the spurions, which contain the low frequency parts of that field,
unchanged. The effect is that the Klaiber regulator, $\kappa$, does not
disappear from physical matrix elements and the solution is no longer
translationally invariant even in the physical subspace. To cure that
problem we must modify the spurions in addition to making the gauge
transformation specified by ${\Theta}$. The correct solution in light-cone
gauge is then given by (we remove the tilde from the ${\eta}$ field since we
keep only the $x^+$ dependent part): 
\begin{equation}
{\Psi}_+ = Z_+ {\rm e}^{{\Lambda}_+^{(-)}}{\sigma}_+ {\rm e}^
{{\Lambda}_+^{(+)}}  
\label{eq:(2.2)}
\end{equation}
\begin{equation}
{\Lambda}_+ = -2i{\sqrt{\pi}}({\eta}(x^+) + \tilde{\Sigma}(x^+,x^-))
\end{equation}
\begin{equation}
Z_+^2 = {\frac{m^2e^\gamma}{8\pi\kappa}} \label{eq:(2.4)}
\end{equation}
\begin{equation}
{\Psi}_- = Z_-{\rm e}^{{\Lambda}_-^{(-)}}{\sigma}_- {\rm e}^{{\Lambda}_-^{(+)}}  
\label{eq:(2.5)}
\end{equation}
\begin{equation}
\Lambda_- = -2i\sqrt{\pi}\phi(x^+)
\end{equation}
\begin{equation}
Z_-^2 = {\frac{\kappa {\rm e}^\gamma}{2\pi}}  \label{eq:(2.7)}
\end{equation}
\begin{equation}
A_+ = {\frac{2}{m}} \partial_+ (\eta + \tilde{\Sigma}) \label{eq:(2.8)}
\end{equation}
\begin{equation}
A_- = 0
\end{equation}
\begin{equation}
\sigma_+ ={\rm exp}\Big[ i\sqrt{\pi}\{Q_5 + Q\}(4m)^{-1} +\int_{0}^\kappa dk_1
k_1^{-1}\{\eta(k_1)-\eta^*(k_1)\}\Big]  \label{sp}
\end{equation}
\begin{equation}
\sigma_- ={\rm exp}\Big[ i\sqrt{\pi}\{Q_5 - Q\}(4m)^{-1} + \int_0^{\kappa} dk_1
k_0^{-1}\{c(k_1) - c^*(k_1)\}\Big]  \label{sm}
\end{equation}
\begin{equation}
Q= \int_{-{\infty}}^{\infty}m{\partial}_+({\phi}-{\eta})dx^+
\end{equation}
\begin{equation}
Q_5= \int_{-{\infty}}^{\infty}m{\partial}_+({\phi}+{\eta})dx^+.
\end{equation}
Again, the vacuum must be chosen to be of the form:
\begin{equation}
|\Omega (\theta )\rangle \equiv \sum_{n=-\infty }^{\infty }{\rm e}^{iM\theta
}|\Omega (M)\rangle \quad ;\quad |\Omega (M)\rangle =({\sigma }_{+}^{\ast }{%
\sigma }_{-})^{M}|0\rangle.
\end{equation}
Now we see that the field, $\Psi_-$, is isomorphic to the left-moving
component of a free, massless Fermi field and it has no dependence on the
ghost field even through the spurion; both of these properties are to be
expected in light-cone gauge$^{3)}$.  Even with the modifications of the
spurions the physics contained in the light-cone gauge solution is the same as
that in the Landau gauge solution.  In particular, we find the anomaly:
\begin{equation}
  \partial^\mu J^5_\mu = {\frac{e}{2\pi}}\varepsilon_{\mu\nu}F^{\mu\nu} \label{avc}
\end{equation}
and the chiral condensate:
\begin{equation}
   \langle\Omega (\theta )|\bar{\Psi}\Psi|\Omega (\theta )\rangle = -{\frac{m}{2\pi}}
{\rm e}^\gamma \cos\theta
\end{equation}

We believe that this construction is the correct light-cone gauge solution
to the continuum Schwinger model but it does have some unexpected properties
which we should discuss. Most striking is that the vacuum expectation of the
spurion, $\sigma_+$, not only does not vanish, as it does in the Landau
gauge solution, but diverges. That fact may cause one to wonder in what
sense the equations of motion are satisfied in the physical subspace. The
point is that the $\Psi_+$ field is not a physical operator ( since it
carries a charge) and the only way the spurions enter physical operators is
in the chargeless combinations, $\sigma_+^*\sigma_-$ and $\sigma_-^*\sigma_+$%
. The chargeless combinations of spurions simply add zero norm states to the
state acted upon and, in particular, act as c-numbers in the physical
subspace. With that in mind, it is easy to use the arguments in ref. 1 to
show that the equations of motion are satisfied in the following sense: Take
any physical operator and use the Lagrange equations of motion to derive an
equation of motion for the physical operator. Then the derived equation of
motion will be satisfied in the physical subspace. That is the common
situation in QCD.

As with the Landau gauge solution, the light-cone gauge solution is
straightforward to quantized on $t = 0$. Indeed, except for the spurions, it
is very similar to the periodic solution found by Bassetto, Nardelli and
Vianello.$^{4)}$ But it is not so straightforward to quantize the continuum
solution on $x^+ = 0$ due to the fact that the current, $\Psi_+^*\Psi_+$ is
not regulated by splitting in the $x^-$ direction. We notice that the
problem was already present in the Landau gauge solution and is due to the
initial value surface, not the gauge choice.

\section{Quantization in the light-cone representation}

The Schwinger model is defined by the Lagrangian: 
\begin{eqnarray}
&L=-{\frac{1}{4}}F_{{\mu }{\nu }}F^{{\mu }{\nu }}-{\lambda }(A_{0}-A_{1})+i%
\bar{\Psi}{\gamma }^{\mu }{\partial }_{\mu }{\Psi }-e\bar{\Psi}{\gamma }%
^{\mu }{\Psi }A_{\mu }=2F_{+-}F_{+-}&  \nonumber \\
&-2{\lambda }A_{-}+2i{\Psi }_{-}^{\ast }{\partial }_{-}\Psi _{-}+2i{\Psi }%
_{+}^{\ast }{\partial }_{+}\Psi _{+}-2e{\Psi }_{-}^{\ast }\Psi _{-}A_{-}-2e{%
\Psi }_{+}^{\ast }\Psi _{+}A_{+}&  \label{eq:(3.1)}
\end{eqnarray}
where: 
\begin{equation}
A_{\pm }={\frac{1}{2}}(A_{0}{\pm }A_{1}),\quad F_{+-}={\partial }_{+}A_{-}-{%
\partial }_{-}A_{+}.  \label{eq:(3.2)}
\end{equation}
The field equations and the gauge fixing condition are: 
\begin{eqnarray}
&2{\partial }_{-}F_{-+}=-e\Psi _{+}^{\ast }\Psi _{+}=-J_{-},&
\label{eq:(3.3)} \\
&2{\partial }_{+}F_{-+}-{\lambda }=e\Psi _{-}^{\ast }\Psi _{-}=J_{+},&
\label{eq:(3.4)} \\
&i{\partial }_{-}\Psi _{-}=e\Psi _{-}A_{-},&  \label{eq:(3.5)} \\
&i{\partial }_{+}\Psi _{+}=e\Psi _{+}A_{+},&  \label{eq:(3.6)} \\
&A_{-}=0.&  \label{eq:(3.7)}
\end{eqnarray}
From current conservation, ${\partial }_{+}J_{-}+{\partial }_{-}J_{+}=0$,
and Eqs.(\ref{eq:(3.3)})and (\ref{eq:(3.4)}), we obtain the field equation
of ${\lambda }$: 
\begin{equation}
{\partial }_{-}{\lambda }=0.  \label{eq:(3.8)}
\end{equation}
The canonical energy-momentum tensor is given by: 
\begin{equation}
T_{{\mu }{\nu }}=i\bar{\Psi}{\gamma }_{\nu }{\partial }_{\mu }{\Psi }-F_{{%
\nu }{\rho }}{\partial }_{\mu }A^{\rho }+g_{{\mu }{\nu }}\{{\frac{1}{4}}F_{{%
\rho }{\sigma }}F^{{\rho }{\sigma }}+{\lambda }(A_{0}-A_{1})\}
\label{eq:(3.9)}
\end{equation}
where we have used the field equation of the Fermion field. Components in
light-cone coordinates are given explicitly by : 
\begin{eqnarray}
&T_{++}=i\Psi _{-}^{\ast }{\partial }_{+}\Psi _{-}-(F_{+-},{\partial }
_{+}A_{+})_{+}&  \nonumber \\
&=i\Psi _{-}^{\ast }{\partial }_{+}\Psi _{-}+({\partial }_{-}A_{+},{\partial 
} _{+}A_{+})_{+}, &  \label{eq:(3.10)} \\
&T_{-+}=i\Psi _{-}^{\ast }{\partial }_{-}\Psi _{-}-(F_{+-})^{2}-(F_{+-},{\
\partial }_{-}A_{+})_{+}&  \nonumber \\
&=i\Psi _{-}^{\ast }{\partial }_{-}\Psi _{-}+({\partial }_{-}A_{+})^{2}, &
\label{eq:(3.11)} \\
&T_{+-}=i\Psi _{+}^{\ast }{\partial }_{+}\Psi _{+}-(F_{+-})^{2}-(F_{+-},{\
\partial }_{+}A_{-})_{+}&  \nonumber \\
&=i\Psi _{+}^{\ast }{\partial }_{+}\Psi _{+}-({\partial }_{-}A_{+})^{2}, &
\label{eq:(3.12)} \\
&T_{--}=i\Psi _{+}^{\ast }{\partial }_{-}\Psi _{+}+(F_{+-},{\partial }%
_{-}A_{-})_{+}&  \nonumber \\
&=i\Psi _{+}^{\ast }{\partial }_{-}\Psi _{+} &  \label{eq:(3.13)}
\end{eqnarray}
where we have used the gauge fixing condition.

From these expressions we can see some of the problems we will encounter when 
we apply the canonical formulation to the Schwinger model in the representation generated by modes of the fields along either light-cone surface. When we 
construct a light-cone-temporal gauge formulation,$^{5)}$ in which 
$x^-$ is chosen to be the evolution parameter, we use $T_{++}$ and 
$T_{-+}$ as densities to calculate the translational generators. We see that ${\Psi}_+$ and ${\Psi}_+^{\ast}$ are not contained in the densities so that we 
can not treat ${\Psi}_+$ as a degree of freedom.  If we consider the standard 
light-cone gauge treatment (the light-cone-axial gauge), in which $x^+$ is the evolution parameter and $T_{+-}$ and $T_{--}$ are the densities of the 
translational generators we see that we need zero mode fields (fields which are functions only of $x^+$).  These fields will require special treatment. 

We shall show in section 4 that we can find a light-cone-temporal gauge solution
by expressing ${\Psi}_+$ as a functional of $A_+$ (it is done the other way around in the light-cone gauge). The problem of the zero-mode fields is principally one of recognizing them. In fact we can find those missing terms by defining the translational generators in the light-cone coordinate space by requiring that they are identical to those in ordinary coordinate space($x^{0},x^{1}$).$^{6)}$ From the divergence equation 
\begin{equation}
{\partial }^{\nu }T_{{\mu}{\nu }}=0  \label{eq:(3.14)}
\end{equation}
we obtain: 
\begin{equation}
{\oint }T_{{\mu}{\nu }}d{\sigma }^{\nu }=0.  \label{eq:(3.15)}
\end{equation}
If we perform the integral over the closed surface shown in Fig.1, it is
clear that the integral over the surface $t=x^{0}$ is the negative of that
over the light-cone surfaces. Thus we obtain 
\begin{eqnarray*}
&{\int }_{-L}^{L}T_{{\mu}0}(x^{0},x^{1})dx^{1}={\int }%
_{x^{0}-L}^{x^{0}+L}T_{{\mu}+}(x^{+},x^{-}=x^{0})dx^{+}& \\
&+{\int }_{x^{0}-L}^{x^{0}}T_{{\mu}-}(x^{+}=x^{0}+L,x^{-})dx^{-}+{\int }%
_{x^{0}}^{x^{0}+L}T_{{\mu}-}(x^{+}=x^{0}-L,x^{-})dx^{-}.& \\
&&
\end{eqnarray*}
Hence in the limit $L{\rightarrow }{\infty }$ we obtain 
\begin{eqnarray}
&P_{\mu}=\int_{-{\infty }}^{\infty }T_{{\mu}0}(x^{0},x^{1})dx^1={\int }_{-{%
\infty }}^{\infty }T_{{\mu}+}(x^+,x^-=x^0)dx^+&  \nonumber \\
&+{\int }_{-{\infty }}^{x^{0}}T_{{\mu}-}(x^+={\infty },x^-)dx^-+{\int }%
_{x^{0}}^{\infty }T_{{\mu}-}(x^+=-{\infty },x^-)dx^-.&  \label{eq:(3.16)}
\end{eqnarray}
We show in section 4 that $T_{{\mu}-}$ is expressed solely in terms of a 
massive field so that $T_{{\mu}-}$ tends to 0 when $x^{+}\to
{\pm }{\infty}$. In that case we have: 
\begin{equation}
P_+=\int_{-{\infty }}^{\infty}\{ i{\Psi _-}^{\ast }{\partial}_+{\Psi }_-
+({\partial }_-A_+,{\partial}_+A_+)_+ \}dx^+, \label{eq:(3.17)}
\end{equation}
\begin{equation}
P_-=\int_{-{\infty }}^{\infty}({\partial}_-A_+)^2 dx^+. \label{eq:(3.18)}
\end{equation}
We see from this that in the temporal gauge formulation there are no missing 
degrees of freedom. We also see that the canonical momenta of $A_{+}$ and 
$\Psi _{-}$ are $2{\partial}_{-}A_{+}$ and $i\Psi _{-}^{\ast }$ respectively. 

\begin{figure}
\centerline{
\psfig{figure=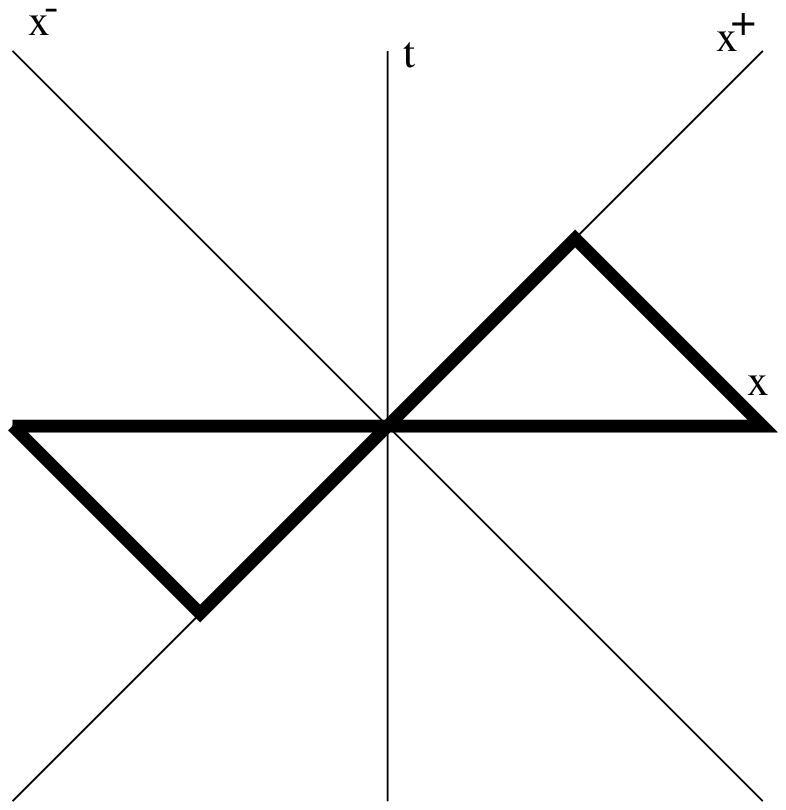,width=3.3in,height=4.3in}
}
\caption{}
\end{figure}

Similarly carrying out the contour integral shown in Fig.2, we obtain: 
\begin{eqnarray}
&P_{\mu}=\int_{-{\infty }}^{\infty }T_{{\mu}0}(x^{0},x^{1})dx^1={\int }_{-{%
\infty }}^{\infty }T_{{\mu}-}(x^+=x^0,x^-)dx^-&  \nonumber \\
&+{\int }_{-{\infty }}^{x^{0}}T_{{\mu}+}(x^+,x^-={\infty })dx^++{\int }%
_{x^{0}}^{\infty }T_{{\mu}+}(x^+,x^-=-{\infty })dx^+.&  \label{eq:(3.19)}
\end{eqnarray}
We show in section 4 that if we choose nonvanishing initial values, then 
$T_{++}$ tends to them in the limit $x^-\to{\pm }{\infty }$. In that case 
we have
\begin{equation}
P_+=\int_{-{\infty }}^{\infty}\{ i{\Psi}_+^{\ast }{\partial}_+{\Psi}_+
-({\partial}_-A_+)^2 \}dx^- +\int_{-{\infty}}^{\infty}T_{++}(
x^+,x^-={\pm}{\infty})dx^+. \label{eq:(3.20)}
\end{equation}
We also show that $T_{-+}$ vanishes in the limit $x^-\to{\pm }{\infty }$
so that we have: 
\begin{equation}
P_-=\int_{-{\infty }}^{\infty }T_{--}(x^+,x^-)dx^-
=\int_{-{\infty }}^{\infty }i{\Psi}_+^{\ast }{\partial}_-{\Psi}_+dx^-. 
  \label{eq:(3.21)}
\end{equation}

\begin{figure}
\centerline{
\psfig{figure=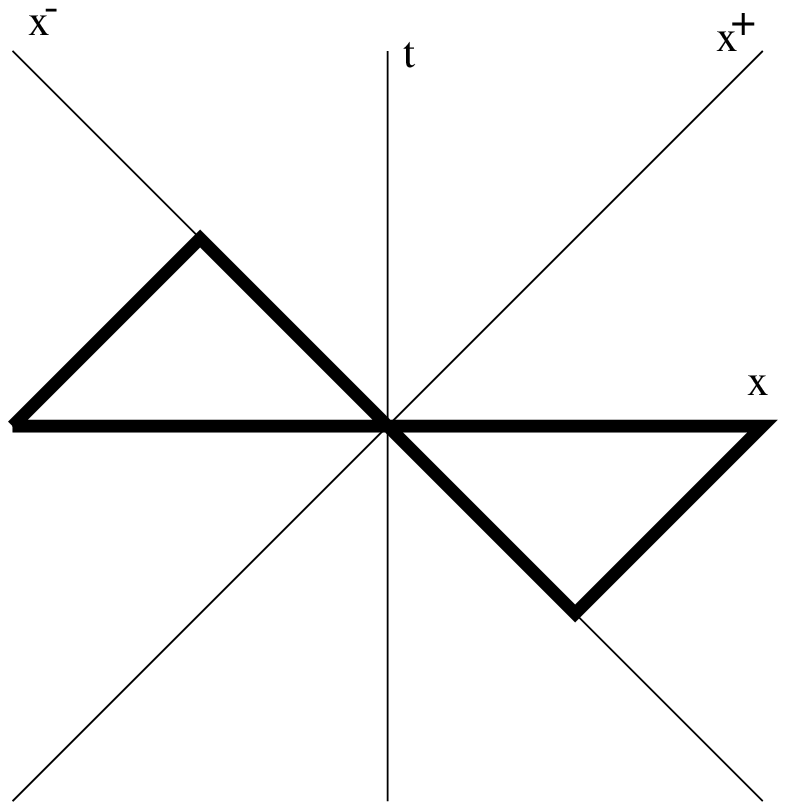,width=3.3in,height=4.3in}
}
\caption{}
\end{figure}

Now we derive quantization conditions for the canonical fields $A_{+}$, ${\
\Psi }_{-}$ and $\Psi _{+}$ by requiring that their commutation relations
with $P_{+}$ and $P_{-}$ give rise to Heisenberg equations which are
consistent with the field equations (the argument is similar to that of
6)). In the temporal gauge formulation $P_+$ in (\ref{eq:(3.17)}) is the 
kinematical operator so that we obtain the Heisenberg equation: 
\begin{equation}
i[P_{+},\Psi _{-}(x)]={\partial }_{+}{\Psi }_-(x)  \label{eq:(3.22)}
\end{equation}
if we require the equal-$x^{-}$ quantization conditions: 
\begin{eqnarray}  \label{eq:(3.23)}
&\{\Psi _{-}(x^{+},x^{-}),\Psi _{-}^{\ast }(y^{+},x^{-})\}_{+}={\delta }%
(x^{+}-y^{+}),&  \nonumber \\
&\{\Psi _{-}(x^{+},x^{-}),\Psi _{-}(y^{+},x^{-})\}_{+}=0,& 
\end{eqnarray}
\begin{eqnarray}
&\lbrack A_{+}(x^{+},x^{-}),\Psi _{-}(y^{+},x^{-})]=0,&  \nonumber \\
&[{\partial }_{-}A_{+}(x^{+},x^{-}),\Psi _{-}(y^{+},x^{-})]=0. &
\label{eq:(3.24)}
\end{eqnarray}
Furthermore, we obtain the Heisenberg equations: 
\begin{equation}
i[P_{+},A_{+}(x)]={\partial }_{+}A_{+}(x),\quad i[P_{+},\Psi _{+}(x)]={%
\partial }_{+}\Psi _{+}(x)  \label{eq:(3.25)}
\end{equation}
if we require, in addition, the equal-$x^{-}$ quantization conditions: 
\begin{eqnarray}
&\lbrack A_{+}(x^{+},x^{-}),A_{+}(y^{+},x^{-})]=0,&  \nonumber \\
&[{\partial }_{-}A_{+}(x^{+},x^{-}),A_{+}(y^{+},x^{-})]=-{\frac{i}{2}}{%
\delta } (x^{+}-y^{+}).&  \label{eq:(3.26)}
\end{eqnarray}
\begin{equation}
\{\Psi _{+}(x^{+},x^{-}),\Psi _{-}^{\ast }(y^{+},x^{-})\}_{+}=0,\quad \{{\Psi }%
_{+}(x^{+},x^{-}),\Psi _{-}(y^{+},x^{-})\}_{+}=0,  \label{eq:(3.27)}
\end{equation}
\begin{eqnarray}
&\lbrack A_{+}(x^{+},x^{-}),\Psi _{+}(y^{+},x^{-})]=0,&  \nonumber \\
&[{\partial }_{-}A_{+}(x^{+},x^{-}),\Psi _{+}(y^{+},x^{-})]=\frac{e}{4}
{\epsilon }(x^{+}-y^{+})\Psi _{+}(x).&  \label{eq:(3.28)}
\end{eqnarray}
We remark that the second commutation relation in (\ref{eq:(3.28)}) is unusual 
in the canonical formalism and that at this point, nothing is known about the equal-$x^-$ commutation relations between ${\Psi}_+$ and ${\Psi}_+^{\ast}$. 
We also remark  that, although $A_+$ obeys a field equation of the 
second order in the light-cone temporal gauge formulation , as is seen from 
(\ref{eq:(3.3)}), the commutator $[{\partial}_-A_+(x^+,x^-),
{\partial}_-A_+(y^+,x^-)]$ is not zero but has the following nonvanishing value
\begin{equation}
[{\partial}_-A_+(x^+,x^-),{\partial}_-A_+(y^+,x^-)]=-i\frac{m^2}{16}
{\epsilon}(x^+-y^+).  \label{eq:(3.29)}
\end{equation}
This is because consistent operator solutions are obtained if and only if we regularize the Fermi products in a gauge invariant way. (In the Schwinger model, regularizing the current operators and the Fermionic kinetic terms 
gauge invariantly gives rise to the chiral anomaly.) It is shown in section 4 that (\ref{eq:(3.23)}) combined with gauge invariant point splitting for the term $i{\Psi}_-^{\ast}{\partial}_+
{\Psi}_-$ gives rise to $-\frac{m^2}{4}A_+^2$ so that (\ref{eq:(3.29)}) 
is required to produce the Heisenberg equation 
\begin{equation}
i[P_+,{\partial}_-A_+(y^+,x^-)]={\partial}_+{\partial}_-A_+(x). 
\label{eq:(3.30)}
\end{equation}

Now that we have obtained the quantization conditions in the temporal gauge 
formulation, we can make use of them to obtain the Heisenberg equations 
which the dynamical $P_-$ in (\ref{eq:(3.18)}) produces. Straightforward 
calculation gives   
\begin{eqnarray}
&i[P_-,{\Psi}_-(x)]=0,\quad i[P_-,A_+(x)]={\partial}_-A_+(x),& \nonumber \\
&i[P_-,{\partial}_-A_+(x)]=-\frac{m^2}{4}({\partial}_+)^{-1}{\partial}_-A_+(x)
,& \nonumber \\
& i[P_-,{\Psi}_+(x)]=-\frac{ie}{2}(({\partial}_+)^{-1}{\partial}_-A_+(x),
{\Psi}_+(x))_+.& \label{eq:(3.31)}
\end{eqnarray} 
We see from the Heisenberg equation of ${\partial}_-A_+$ that ${\partial}_-A_+$
 behaves like a free field of mass $m$. 

In the axial gauge formulation $P_-$ in (\ref{eq:(3.21)}) is the kinematical 
operator so that we obtain the Heisenberg equations: 
\begin{equation}
i[P_{-},\Psi _{-}(x)]=0,\quad i[P_{-},\Psi _{+}(x)]={\partial }_{-}\Psi
_{+}(x)  
\end{equation}
if we specify the following equal-$x^{+}$ commutation relations: 
\begin{equation}
\{\Psi _{-}(x^{+},x^{-}),\Psi _{+}^{\ast }(x^{+},y^{-})\}_{+}=0,\quad 
\{\Psi_{-}(x^{+},x^{-}),{\Psi }_{+}(x^{+},y^{-})\}_{+}=0,  \label{eq:(3.33)}
\end{equation}
\begin{eqnarray}
&\{\Psi _{+}(x^{+},x^{-}),\Psi _{+}^{\ast }(x^{+},y^{-})\}_{+}={\delta }
(x^{-}-y^{-}),&  \nonumber \\
&\{{\Psi }_{+}(x^{+},x^{-}),\Psi _{+}(x^{+},y^{-})\}_{+}=0.&
\label{eq:(3.34)}
\end{eqnarray}
We can not obtain any other quantization conditions unless we solve 
Eqs.(\ref{eq:(3.3)}) and (\ref{eq:(3.6)}). 

\section{Construction of operator solutions}

Now that we have the algebra of the fields, we can proceed to construct the
solution. \ We might consider the problem as an initial value problem on $%
x^-=0$ or $x^+=0$, or, proceed in a more covariant way. \ Here, we shall
consider the initial value problem on each characteristic. \ First we consider 
the initial value problem on the surface $x^-=0$. 

\subsection{Light-cone temporal gauge solution}
From (\ref{eq:(3.5)}) we see 
that when $A_{-}$=0, the first component, $\Psi_{-}$, is a free field depending only on $x^{+}$. Thus we specify $\Psi _{-}$ as a free, massless Fermion field: \begin{equation}
\Psi _{-}(x)=\psi _{-}(x)  \label{eq:(4.1)}
\end{equation}
satisfying the anticommutation relations: 
\begin{equation}
\{\psi _{-}(x),{\psi }_{-}^{\ast }(y)\}_{+}={\delta }(x^{+}-y^{+}),
\quad \{\psi_{-}(x),{\psi }_{-}(y)\}_{+}=0.  
\end{equation}
Furthermore we make use of the fusion field defined by: 
\begin{equation}
:e{\psi }_{-}^{\ast }(x)\psi _{-}(x):=m{\partial }_{+}{\phi }(x)
\label{eq:(4.3)}
\end{equation}
to express $\psi _{-}$ in the following equivalent bosonized form: 
\begin{equation}
\psi _{-}(x)=Z_{-}{\rm exp}[-2i\sqrt{\pi }{\phi }^{(-)}(x)]{\sigma }_{-}{\rm %
exp}[-2i\sqrt{\pi }{\phi}^{(+)}(x)].  \label{eq:(4.4)}
\end{equation}
Here, $Z_{-}$ is the finite normalization constant , ${\sigma }_{-}$ is the
spurion operator and ${\phi }^{(-)}$ and ${\phi }^{(+)}$ are positive and
negative frequency parts of ${\phi }$ regularized {\it a la} Klaiber (the
construction is exactly the same as in (\ref{eq:(2.5)}),(\ref{eq:(2.7)}) and 
(\ref{sm})). By construction ${\phi}$ satisfies the following commutation 
relation: 
\begin{equation}
\lbrack {\phi }(x),{\phi }(y)]=-{\frac{i}{4}}{\epsilon }(x^{+}-y^{+}).
\label{eq:(4.5)}
\end{equation}
Then, carrying out a gauge invariant point splitting procedure we find the
following well-defined current: 
\begin{eqnarray}
&J_{+}(x)=\lim_{y^{+}\rightarrow x^{+}}{\frac{e}{2}}\{{\psi }_{-}^{\ast
}(x^{+})\psi _{-}(y^{+}){\rm exp}[-ie%
\int_{y^{+}}^{x^{+}}A_{+}(z^+,x^-)dz^+]+h.c.\}&  \nonumber \\
&=m{\partial }_{+}{\phi }(x)-{\frac{m^{2}}{2}}A_{+}(x). &  \label{eq:(4.6)}
\end{eqnarray}
Furthermore, the kinetic term $i{\Psi}_-^{\ast}{\partial}_+{\Psi}_-$ is 
regularized as follows: 
\begin{eqnarray}
&\lim_{y^{+}\rightarrow x^{+}}\{{\frac{i}{2}}{\psi}_{-}^{\ast}(x^+){%
\partial }_{+}\psi _{-}(y^{+}){\rm exp}[-ie%
\int_{y^{+}}^{x^{+}}A_{+}(z,x^{-})dz]+h.c.-{\frac{1}{2{\pi }}}{\frac{1}{%
(x^{+}-y^{+})^{2}}}\}&  \nonumber \\
&=({\partial }_{+}{\phi })^{2}-{\frac{m^{2}}{4}}(A_{+})^{2}. &
\label{eq:(4.7)}
\end{eqnarray}
so that $P_+$ in (\ref{eq:(3.17)}) is given by  
\begin{equation}
P_+={\int }_{-{\infty }}^{\infty }\{({\partial }_{+}{\phi })^{2}-{\frac{m^{2}}
{4}}(A_{+})^{2}+({\partial }_{-}A_{+},{\partial }_+A_+)_+ \}dx^+. 
\label{eq:(4.8)}
\end{equation}

Now note that owing to (\ref{eq:(4.6)}), Eq.(\ref{eq:(3.4)}) can be written as  
\begin{equation}
(4{\partial }_{+}{\partial }_{-}+m^{2})A_{+}(x)=2\{{\lambda }(x)+m{\partial }%
_{+}{\phi }(x)\}.  \label{eq:(4.9)}
\end{equation}
Multiplying this by ${\partial}_-$ leads to
\begin{equation}
(4{\partial }_+{\partial }_-+m^2){\partial}_-A_+(x)=0  \label{eq:(4.10)}
\end{equation}
due to the fact that  ${\phi }(x)$ and ${\lambda }(x)$ depend only on $x^{+}$. 
Thus we see that ${\partial}_-A_+$ is a free field of mass $m$. Since 
${\partial}_-A_+$ is gauge invariant and is equal to
 $-\frac{m}{2}\tilde{\Sigma}$ in the Landau gauge operator solution, 
our present result is in agreement with the earlier results.  Here we set
\begin{equation}
{\partial }_-A_+(x)=-\frac{m}{2}\tilde{\Sigma}.  \label{eq:(4.11)}
\end{equation}
where the normalization is determined by (\ref{eq:(3.29)}). Then we notice 
that replacing ${\partial}_-A_+$ in (\ref{eq:(4.9)}) 
by $\tilde{\Sigma}$ enables us to express $A_+$ in terms of 
${\phi},{\lambda}$ and $\tilde{\Sigma}$ as
\begin{equation}
A_+(x)=\frac{2}{m^2}\{ {\lambda}+m{\partial}_+({\phi}+\tilde{\Sigma}) \}
  \label{eq:(4.12)}.
\end{equation}
The commutation relation (\ref{eq:(3.29)}) is transcribed as that of 
$\tilde{\Sigma}$:
\begin{equation}
[\tilde{{\Sigma}}(x^+,x^-),\tilde{{\Sigma}}(y^+,x^-)] =-{\frac{i}{4}}
{\epsilon}(x^+-y^+)  \label{eq:(4.13)}
\end{equation}
and the commutation relations of ${\lambda}$ are obtained by rewriting 
(\ref{eq:(4.9)}) for ${\lambda}$ as
\begin{equation}
{\lambda}=\frac{1}{2}\{ m^2A_++4{\partial}_+({\partial}_-A_+)
-m{\partial}_+{\phi} \}   \label{eq:(4.14)}
\end{equation}
and by making use of the commutation relations (\ref{eq:(3.26)}), 
(\ref{eq:(3.29)}),(\ref{eq:(4.5)}) and 
\begin{equation}
\lbrack {\phi }(x),A_{+}(y)]=0,\quad [{\phi }(x),{\partial }_{-}A_{+}(y)]=0,
 \label{eq:(4.15)}
\end{equation}
which result from (\ref{eq:(3.24)}),(\ref{eq:(4.1)}) and (\ref{eq:(4.3)}). 
Combining these results we get
\begin{equation}
[{\lambda}(x),{\lambda}(y)]=0, \quad [{\lambda}(x),\tilde{\Sigma}(y)]=0, \quad
[{\lambda}(x),{\phi}(y)]=i\frac{m}{2}{\delta}(x^+-y^+). \label{eq:(4.16)}
\end{equation}

Now we see that ${\lambda}$ is a zero norm field and that Maxwell's equations 
are recovered in a physical subspace formed by factoring the zero norm field 
${\lambda}$ out of the representation space. 
If we rewrite ${\lambda}$ as 
\begin{equation}
{\lambda }(x)=m{\partial }_{+}({\eta }(x)-{\phi }(x))  \label{eq:(4.17)}
\end{equation}
and take account of the third commutation relation of ${\lambda}$ in(\ref{eq:(4.16)})  
we find that ${\eta}$ is a negative norm field depending only on $x^+$ and satisfies 
the following commutation relations
\begin{equation}
\lbrack {\eta}(x),{\eta}(y)]={\frac{i}{4}}{\epsilon}(x^+-y^+),\quad 
[{\eta}(x),\tilde{{\Sigma}}(y)]=0,\quad [{\eta}(x),{\phi}(y)]=0. 
\label{eq:(4.18)}
\end{equation}
As a result $A_+$ in (\ref{eq:(4.12)}) may be written as
\begin{equation}
A_{+}={\frac{2}{m}}{\partial }_{+}(\tilde{{\Sigma }}+{\eta }).\label{eq:(4.19)}
\end{equation}
In terms of these fields $P_+$ in (\ref{eq:(4.8)}) and $P_-$ in (\ref{eq:(3.18)}) are 
diagonalized as follows 
\begin{equation}
P_+=\int_{-{\infty}}^{\infty}\{ ({\partial}_+{\phi})^2-({\partial}_+{\eta})^2+
({\partial}_+\tilde{\Sigma})^2 \}dx^+,  \label {eq:(4.20)} 
\end{equation}
\begin{equation}
P_-=\int_{-{\infty}}^{\infty}\frac{m^2}{4}(\tilde{\Sigma})^2dx^+, 
\label{eq:(4.21)}
\end{equation}
which shows that ${\phi},{\eta}$ and $\tilde{\Sigma}$ are constituent 
free fields of the light-cone temporal gauge Schwinger model. 

Let us turn to specifying ${\Psi}_+$, which satisfies (\ref{eq:(3.3)}) and 
(\ref{eq:(3.6)}). Because in the temporal gauge formulation there is no 
dynamical equation which allows us to determine ${\Psi}_+$ as an initial value 
problem, we make use of the fact that using (\ref{eq:(4.19)}), we can write Eq.
(\ref{eq:(3.3)}) as 
\begin{equation}
J_-=e{\Psi}_+^{\ast}{\Psi}_+=m{\partial}_-\tilde{\Sigma}.
\end{equation}
We see from this that the $\tilde{\Sigma}$ field can be identified as a 
fusion field composed of ${\Psi}_+^{\ast}$ and ${\Psi}_+$ and that 
${\Psi}_+$ in turn can be expressed in an equivalent bosonized form(this 
result will be obtained in the axial gauge formulation). As a matter 
of fact $A_+$ given in (\ref{eq:(4.19)}) is identical with the electromagnetic 
field given in(\ref{eq:(2.8)}) so that the Fermion operator (\ref{eq:(2.2)}) 
satisfies Eq.(\ref{eq:(3.6)}). Therefore we specify ${\Psi}_+$  as in(\ref{eq:(2.2)}) 
and show that it also satisfies Eq.(\ref{eq:(3.3)}). 
To this end we also regularize the bilinear product 
$e{\Psi}_+^{\ast}{\Psi}_+$ by the gauge-invariant point splitting 
procedure. We see immediately that if we split only in the $x^+$ 
direction, then the sum $\tilde{\Sigma}+{\eta}$ in the exponent behaves like 
a zero norm operator so that the procedure does not work. We also see 
that if we split only in the $x^-$ direction, then the ${\eta}$ field 
gives rise to a divergence at high frequencies. Therefore we have to split 
in another direction. The following two-step limit, with ${\epsilon}$ being 
a timelike vector, gives us the desired result:
\begin{eqnarray}
&J_-(x)=\frac{e}{2}\lim_{{\epsilon}^- \to0} \{ \lim_{{\epsilon}^+ \to0}
\left( {\Psi}_+^{\ast}(x+{\epsilon}){\Psi} _+(x){\rm exp}
[-ie\int_x^{x+{\epsilon}}dz^{\nu}A_{\nu}(z)]+h.c.\right) \}&  \nonumber \\&=m{\partial }_-\tilde{\Sigma}(x). &  \label{eq:(4.23)}
\end{eqnarray}
The axial-vector current $J^5_\mu \equiv \varepsilon_{\mu\nu}J^{\nu}$, where 
${\varepsilon}_{-+}=-{\varepsilon}_{+-}=\frac{1}{2}$ and 
 ${\varepsilon}_{--}={\varepsilon}_{++}=0$, satisfies (\ref{avc}). 
In addition, a conserved axial-vector current 
\begin{equation}
J^5_{C\mu}={\varepsilon}_{{\mu}{\nu}}(J^{\nu}+m^2A^{\nu}), 
\end{equation}
is obtained by regularizing 
$e\bar{\Psi}{\gamma}_{\mu}{\gamma}^5{\Psi}$, 
with ${\Psi}_-$ and ${\Psi}_+$ being ${\psi}_-$  and (\ref{eq:(2.2)}) 
respectively, in the same manner as the vector current:
\begin{eqnarray}
&J_{C+}^5(x)=-\frac{e}{2}\lim_{{\epsilon}^+ \to0}\{{\psi }_-^{\ast}
(x^++{\epsilon}^+)\psi _-(x^+){\rm exp}[ie\int_{x^+}^{x^++{\epsilon}^+}
A_+(z^+,x^-)dz^+]+h.c.\}&  \nonumber \\
&=-\{ m{\partial}_+{\phi}(x)+\frac{m^2}{2}A_+(x) \}
=-m{\partial}_+({\phi}+{\eta}+\tilde{\Sigma}), &  \label{eq:(4.25)}
\end{eqnarray}
\begin{eqnarray}
&J_{C-}^5(x)=\frac{e}{2}\lim_{{\epsilon}^- \to0} \{ \lim_{{\epsilon}^+ \to0}
\left( {\Psi}_+^{\ast}(x+{\epsilon}){\Psi} _+(x){\rm exp}
[ie\int_x^{x+{\epsilon}}dz^{\rho}{\varepsilon}_{{\rho}{\sigma}}A^{\sigma}(z)]+
h.c. \right) \}&  \nonumber \\&=m{\partial }_-\tilde{\Sigma}(x). &  
\label{eq:(4.26)}
\end{eqnarray}
It can be shown furthermore that the Fermion operator (\ref{eq:(2.2)}) 
satisfies the anticommutation relations in (\ref{eq:(3.34)}), if we define 
them as $y^+{\to}x^+$ limit so as to avoid any divergences. 

We end the temporal gauge construction by defining physical space $V$ by
\begin{equation}
V=\{\; |{\rm phys}>\; |\; {\lambda}^{(+)}(x)|{\rm phys}>=0\; \} 
\end{equation}
where ${\lambda}^{(+)}(x)$ denotes the positive frequency part of ${\lambda}$.

\subsection{Light-cone axial gauge solution}
In the axial gauge formulation $x^+$ is taken to be the evolution parameter 
so that we can solve Eq.(\ref{eq:(3.6)}) as an initial value problem on 
$x^+=0$. As an initial value of ${\Psi}_+$ we take a free Fermi field 
${\Psi}_R(x^-)$ and define a fusion field $\tilde{\phi}$ by 
\begin{equation}
:e{\Psi }_{R}^{\ast }(x^-){\Psi }_{R}(x^-):=m{\partial }_{-}\tilde{{%
\phi }}(x^-).  \label{eq:(4.28)}
\end{equation}
Then, by construction $\tilde{\phi}$ satisfies the commutation relation
\begin{equation}
[\tilde{\phi}(x^-),\tilde{\phi}(y^-)]=-\frac{i}{4}{\epsilon}(x^--y^-) 
\end{equation}
and Eq.(\ref{eq:(3.3)}) becomes
\begin{equation}
J_-(0,x^-)=m{\partial}_-\tilde{\phi}(x^-)=-2{\partial}_-^2A_+(0,x^-)
 \label{eq:(4.30)}
\end{equation}
where point splitting in the $x^-$ direction has enabled us to utilize 
the fact that $A_-=0$. From (\ref{eq:(4.30)}) we obtain
\begin{equation}
{\partial}_-A_+(0,x^-)=-\frac{m}{2}\tilde{\phi}(x^-). \label{eq:(4.31)} 
\end{equation} 
As a consequence if we neglect the second unspecified term of (\ref{eq:(3.20)}) 
for the moment, then we can express $P_+$ solely in terms of $\tilde{\phi}$ as
\begin{eqnarray} 
&P_+=\int_{-{\infty}}^{\infty}\{ J_-(0,x^-)A_+(0,x^-)-
({\partial}_-A_+)^2 \}dx^- & \nonumber \\
&= \int_{-{\infty}}^{\infty}({\partial}_-A_+)^2 dx^-=
\frac{m^2}{4}\int_{-{\infty}}^{\infty}(\tilde{\phi}(x^-))^2dx^-. & 
 \label{eq:(4.32)}
\end{eqnarray}
Furthermore, by making use of the equivalent bosonized form of ${\Psi}_R$ 
we can express$P_-$ in (\ref{eq:(3.21)}) as
\begin{equation} 
P_-=\int_{-{\infty}}^{\infty}i{\Psi}_+^{\ast}(0,x^-){\partial}_-
{\Psi}_+(0,x^-)dx^-
= \int_{-{\infty}}^{\infty}({\partial}_-\tilde{\phi})^2 dx^-. 
 \label{eq:(4.33)}
\end{equation}
It follows from (\ref{eq:(4.32)}) and (\ref{eq:(4.33)}) that the fusion 
field $\tilde{\phi}$ is again constituent free field of mass $m$. 

The temporal evolution of ${\Psi}_+(0,x^-)$ is defined by making use of 
$P_+$ in (\ref{eq:(4.32)}) by
\begin{equation}
{\Psi}_+(x^+,x^-){\equiv}{\rm e}^{iP_+x^+}{\Psi}_R(0,x^-){\rm e}^{-iP_+x^+}. 
\end{equation}
Then by making use of the equivalent bosonized form we can write
\begin{equation}
{\Psi}_+(x^+,x^-)=Z{\rm exp}[-2i\sqrt{\pi}\tilde{\phi}^{(-)}(x)]
{\sigma}_R{\rm exp}[-2i\sqrt{\pi}\tilde{\phi}^{(+)}(x)] \label{eq:(4.35)} 
\end{equation}
where  
\begin{equation}
Z^2=\frac{{\tilde{\kappa}}{\rm e}^{\gamma}}{2{\pi}}, \label{eq:(4.36)}
\end{equation}
\begin{equation}
\tilde{\phi}^{(+)}(x)=\frac{i}{\sqrt{4{\pi}}}\int_0^{\infty}
\frac{dp_-}{p_-}c(p_-)({\rm e}^{-ip{\cdot}x}-{\theta}({\tilde{\kappa}}-p_-)), 
\quad \tilde{\phi}^{(-)}(x)=(\tilde{\phi}^{(+)}(x))^{\ast} 
\end{equation}
with $p_+=\frac{m^2}{4p_-}$ and ${\sigma}_R$ is the spurion operator
\begin{equation}
{\sigma}_R={\rm exp}[\int_0^{\tilde{\kappa}}\frac{dp_-}{p_-}
( c(p_-)-c^{\ast}(p_-) )]. \label{eq:(4.38)}
\end{equation}
Here we have omitted the Klein transformation factor. It is evident that 
${\Psi}_+$ satisfies the anticommutation relations (\ref{eq:(3.34)}). 
 
Now we notice that on the surface $x^+=$constant, ${\Psi}_+$ behaves like a 
free fermion field and that both $P_+$ and $P_-$ are diagonalized in terms of 
the fusion field. Thus we see that the common hope in the light-cone 
quantization that the light-cone bare states are closer to partons than the 
ordinary equal-time bare states is realized in its strongest possible form. 
At the same time we also see that the initial value problem which we have 
considered also gives rise to the well-known problem common to axial gauge 
quantizations: we come have ill-defined equal-$x^+$ commutation 
relations of $A_+$ because $A_+$ is obtained from (\ref{eq:(4.31)}) as  
\begin{equation}
A_+=-\frac{m}{2}({\partial}_-)^{-1}\tilde{\phi}=
\frac{2}{m}{\partial}_+\tilde{\phi}. \label{eq:(4.39)}
\end{equation}
Note that the difficulty results from the fact that the antiderivative
$({\partial}_-)^{-1}$ is not well-defined in any positive definite Hilbert 
space.$^{8)}$ Therefore we introduce the ${\eta}$ field as in (\ref{eq:(4.19)}) 
to regularize (\ref{eq:(4.39)}) although doing so obscures the parton picture. 
To obtain a consistent solution we also introduce the 
Fermi field ${\psi}_-(x^+)$ as well as the fusion field ${\phi}(x^+)$ and solve Eqs.(\ref{eq:(3.4)}) and (\ref{eq:(3.5)}) in the same manner as in the temporal 
gauge formulation. This enables us to identify the $\tilde {\Sigma}$ field 
with the fusion field $\tilde{\phi}$ and thus hereafter we denote $\tilde{\phi}$ 
as $\tilde{\Sigma}$ . Furthermore by assuming that the massive degrees 
of freedom of $T_{++}$ contained in (\ref{eq:(4.20)}) vanishes as $x^-\to
{\pm}{\infty}$, we obtain 
\begin{equation}
T_{++}(x^+,x^-={\pm}{\infty})=({\partial}_+{\phi})^2-({\partial}_+{\eta})^2
\end{equation} 
so that $P_+$ in (\ref{eq:(3.20)}) is fixed to be
\begin{equation}
P_{+}= {\frac{m^2}{4}}{\int }_{-{\infty }}^{\infty }\tilde{\Sigma}^2 dx^{-} +
\int_{-{\infty }}^{\infty }\{ ({\partial}_+{\phi})^2 -({\partial}_+{\eta}%
)^2 \}dx^{+}.  \label{eq:(4.41)}
\end{equation}
In this way we can reconstruct, in the axial gauge formulation, the $P_+$ 
given in (\ref{eq:(4.20)}) .

Now that $A_+$ possesses zero mode fields (fields independent of $x^-$), we 
have to take this fact into account when we solve Eq.(\ref{eq:(3.6)}) as the 
initial value problem on the surface $x^+=0$. As an alternative initial value 
satisfying the equal-$x^+$ anticommutation relations we choose
\begin{equation}
{\Psi}_+(0,x^-)={\rm exp}[-2i\sqrt{\pi}{\eta}(0)]{\Psi}_R(x^-).
\label{eq:(4.42)}
\end{equation}
Note that (\ref{eq:(4.42)}) has a diverging vacuum expectation value, 
which is inevitable as long as we respect the equal-$x^+$ anticommutation 
relations. In fact when we rewrite the exponential function of 
${\eta}(0)$ as the normal product, divergences appear at low frequencies and 
at high frequencies but the divergence at high frequencies is canceled by the 
zero from the ${\Psi}_R$, whereas divergence at low frequencies remains. 

To investigate connection between (\ref{eq:(4.42)}) and (\ref{eq:(2.2)}), 
we rewrite the exponential function of ${\eta}(0)$ as follows
\begin{equation}
{\rm e}^{-2i\sqrt{\pi}{\eta}(0)}={\rm exp}[\frac{1}{2}\int_{\kappa}^{\infty}
\frac{dk_+}{k_+}]{\rm exp}[-2i\sqrt{\pi}{\eta}^{(-)}(0)]{\sigma}_+
{\rm exp}[-2i\sqrt{\pi}{\eta}^{(+)}(0)].  \label{eq:(4.43)}
\end{equation} 
At the same time we rewrite the spurion operator ${\sigma}_R$ as the 
normal product form:
\begin{equation}
{\sigma}_R={\rm exp}[-\frac{1}{2}\int_0^{\tilde{\kappa}}\frac{dp_-}{p_-}]
{\rm exp}[-\int_0^{\tilde{\kappa}}\frac{dp_-}{p_-}c^{\ast}(p_-)]
{\rm exp}[\int_0^{\tilde{\kappa}}\frac{dp_-}{p_-}c(p_-)].  \label{eq:(4.44)}
\end{equation}
Then we see that because
\begin{equation}
\int_0^{\tilde{\kappa}}\frac{dp_-}{p_-}=\int_{\frac{m^2}{4\tilde{\kappa}}}
^{\infty}\frac{dp_+}{p_+} 
\end{equation}
the divergence from the former is canceled by one from the latter 
if we require
\begin{equation}
\tilde{\kappa}=\frac{m^2}{4{\kappa}}. 
\end{equation}
In that case the normalization factor $Z$ in (\ref{eq:(4.36)}) is identical
 with the $Z_+$ in (\ref{eq:(2.4)}) so that (\ref{eq:(4.42)}) agrees exactly 
with (\ref{eq:(2.2)}) at $x^+=0$. 

Now  we notice that the Fermion operator (\ref{eq:(2.2)}) is obtained from 
the initial value (\ref{eq:(4.42)}) with $\tilde{\kappa}=\frac{m^2}{4{\kappa}}$  as a result of the temporal evolution
\begin{equation}
{\Psi}_+(x^+,x^-){\equiv}{\rm e}^{iP_+x^+}{\Psi}_+(0,x^-){\rm e}^{-iP_+x^+} 
\end{equation}
and that the zero mode fields do not prevent us from obtaining  
\begin{equation}
J_-=m{\partial}\tilde{\Sigma}, \quad 
i{\Psi}_+^{\ast}{\partial}_-{\Psi}_+=({\partial}_-\tilde{\Sigma})^2  
\end{equation}
and hence both $P_+$ and $P_-$ are already diagonal so that the parton picture is 
realized even when there exist zero mode fields in the formulation. 
That is the main finding of this paper.

We end this section by pointing out that we can not change the order of 
integrations and differentiations in the evaluation of the commutation relations 
of $A_+$. In fact they are obtained unambiguously if we first evaluate 
the two-dimensional commutator $[{\eta}(x)+\tilde{\Sigma}(x),
{\eta}(y)+\tilde{\Sigma}(y)]=iE(x-y)$ and then differentiate the $iE(x-y)$ 
with respect ${\partial}_+^x$ and ${\partial}_+^y$. It turns out that
\begin{eqnarray}
E(x)&=\frac{1}{2{\pi}}\int_0^{\infty}\frac{dk_+}{k_+}
\{ {\rm sin}k_+x^+-{\rm sin}(k_+x^++\frac{m^2}{4k_+}x^-) \}& \nonumber \\
&=\frac{1}{4}{\epsilon}(x^+)-\frac{{\epsilon}(x^+)+{\epsilon}(x^-)}{4}
J_0(m\sqrt{x^2})&
\end{eqnarray} 
where $J_0$ denotes the Bessel function of order 0. 
 
\section{Concluding remarks}

In this paper we have shown that an operator solution to the light-cone
gauge Schwinger model is obtained by solving the initial value problem on 
$x^-=0$ to specify ${\Psi}_-$ and one on $x^+=0$ to specify ${\Psi}_+$ 
simultaneously.  The solution turns out to be independent of any particular
representation.  We have discussed the formulation and solution of the
problem in the equal-time representation and in the light-cone
representation.  The problem is straightforward to formulate at equal
time, but solving it involves a nontrivial diagonalization of the
Hamiltonian.

If we consider the initial value problem on $x^+ = 0$, there are two
difficulties which we must overcome: we must include the zero-mode
fields; and we must regularize the Fermi products on our initial value
surface.  As for the first problem, a substantial ability to find the
necessary zero-mode fields has developed starting with the work of
Bassetto, Soldati and Nardelli$^{9)}$.  Once the necessary fields
are recognized, putting them into the solution is not difficult. 
Looking at the full solution we can see how the Fermi products are 
regulated when the covariant solution is split in a timelike 
direction.  There are four sources of singularity: there is a 
singularity due to the $\tilde{\Sigma}$ field at high frequency; if 
the splitting is not zero in the $x^-$ direction it gives a factor 
proportional to ${(x^-)}^{-1}$---that singularity is necessary for the 
point splitting procedure to work; there is a singularity due to the 
$\tilde{\Sigma}$ field at low frequencies; if the splitting is not zero in 
the $x^+$ direction it gives a factor proportional to ${(x^+)}^{-1}$ and is 
therefore not regulated by splitting in the $x^-$ direction---that 
singularity is cancelled by a zero from the ghost field. 
There is a potential singularity due to the $\eta$ field at high 
frequencies; actually, it turns out to be a zero; if the splitting is 
not zero in the $x^+$ direction it gives a factor proportional to 
${(x^+)}$ and is the zero which cancels the low frequency singularity 
from the $\tilde{\Sigma}$ field.  There is a low frequency 
singularity from the $\eta$ field---that singularity is absorbed into 
the spurion and gives rise to the infrared states, linear 
combinations of which form the $\theta$ states. 

In DLCQ the $p^+ = 0$ singularity is regulated with periodicity conditions.  
In that case the continuum answer is not recovered for physical matrix
elements, even if the problem is solved exactly at finite L and the
limit  $L\rightarrow \infty$ is then taken.$^{10)}$  It may well be
that for many problems it is necessary to carefully regulate the theory
prior to imposing periodicity conditions.  The DLCQ grid would then be
just a numerical device, not a regulator.  Such procedures have been
suggested in ref.11).  We do not think the techniques necessary
to carry out that procedure are known for all cases but the knowledge
is growing.


\section*{Acknowledgments}

The work of one of us (GMc) was supported by grants from the U.S. Department of Energy.






\end{document}